\documentclass{mem}
\usepackage{natbib}\usepackage{txfonts}\usepackage{balance}
\usepackage{graphicx}
\usepackage[a4paper,breaklinks,dvipdfm]{hyperref}
\idline{75}{282}
\begin{document}
\def\teff{$T\rm_{eff }$}
\def\kms{$\mathrm {km s}^{-1}$}

\title{
The horizontal branch morphology of M31 Globular Clusters.}

   \subtitle{}

\author{
M. \,Bellazzini\inst{1}, A. \,Buzzoni\inst{1},
C. \,Cacciari\inst{1}, L. \, Federici\inst{1}, F. \,Fusi Pecci\inst{1}, S. \,Galleti\inst{1}, \and S. \,Perina\inst{1,2}
          }

  \offprints{M. Bellazzini}

\institute{
Istituto Nazionale di Astrofisica --
Osservatorio Astronomico di Bologna, Via Ranzani 1,
I-40127 Bologna, Italy
\and
Departamento de Astronom\'ia y Astrof\'isica, Pontificia Universidad Cat\'olica de Chile, 7820436 Macul, Santiago, Chile
\\
\email{michele.bellazzini@oabo.inaf.it}
}

\authorrunning{Bellazzini et al.}

\titlerunning{HB Morphology in M31 CGs}

\abstract{
We present the results of a first global analysis of the Horizontal Branch morphology of Globular Clusters in the nearby spiral M31, based on their Color Magnitude Diagrams. 
\keywords{galaxies: individual: M~31 -- Globular Clusters}
}
\maketitle{}

\section{Introduction}

\begin{figure*}[t!]
\resizebox{\hsize}{!}{\includegraphics[clip=true]{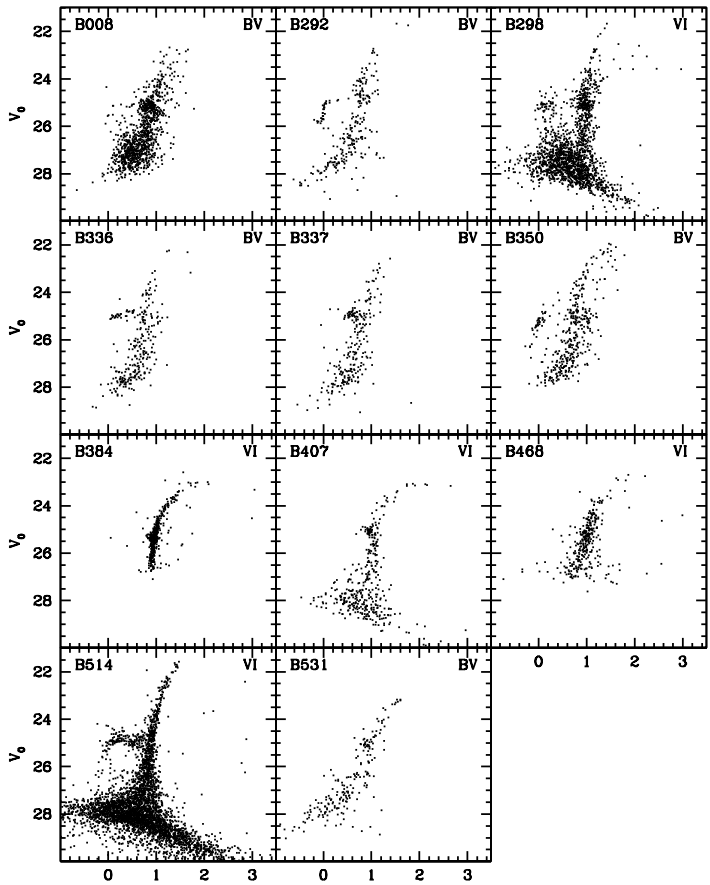}}
\caption{\footnotesize CMDs of eleven clusters in the considered sample.
}
\label{cmd1}
\end{figure*}
\begin{figure*}[t!]
\resizebox{\hsize}{!}{\includegraphics[clip=true]{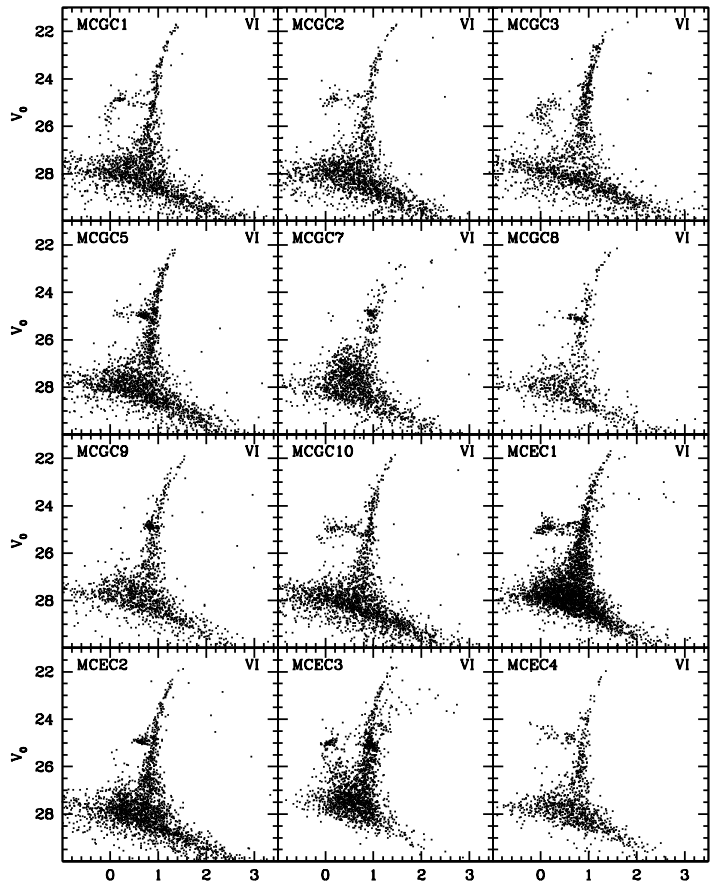}}
\caption{\footnotesize CMDs of the remaining twelve clusters in the sample.
}
\label{cmd2}
\end{figure*}

\begin{figure*}[t!]
\resizebox{\hsize}{!}{\includegraphics[clip=true]{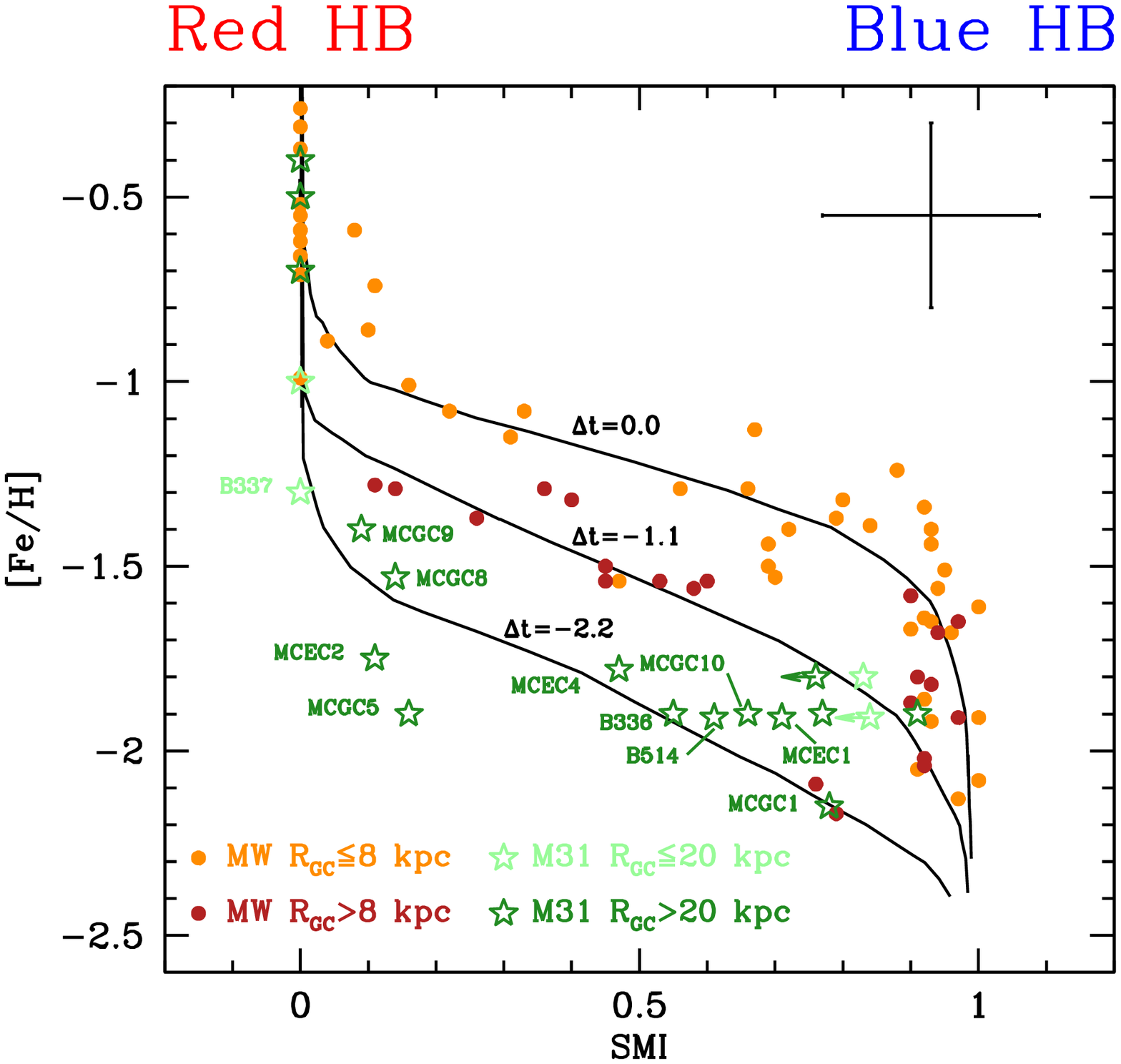}}
\caption{\footnotesize The SMI vs. metallicity diagram. The Galactic GCs are
 plotted as filled circles, the M31 GCs as open stars.  The symbol color code shows the  
 galacto-centric distance of the cluster, as  described in the bottom-left corner of the panel.
 Arrows indicate upper limits. Solid lines are isochrones from the synthetic HB models by
 \citet{rey01}, and are labelled according to their age difference in Gyr. Typical error-bars 
 for M31 GCs are shown in the top-right corner. M31 clusters lying near the $\Delta t=-2.2$~Gyr 
 isochrone are labelled.
}
\label{mir}
\end{figure*}

Deep, high quality Colour Magnitude Diagrams (CMD) from HST observations have been used to compute a simplified version of the Mironov index (SMI\footnote{SMI is defined as $\frac{B}{B+R}$, where B and R are the number of HB stars to the blue or to the red of a given color threshold \citep[see][]{sibi}.}), to parameterize the Horizontal Branch (HB) morphology for 23 globular clusters (GCs) in the M31 galaxy \citep[from the sample presented in][see Fig.~\ref{cmd1} and Fig.~\ref{cmd2}]{papI}. All the considered clusters are located in the outer halo at projected distances between 10 and 100 kpc from the galaxy center. This allowed us to probe the SMI vs. [Fe/H] relationship (Fig.~\ref{mir}) by comparing them with their Galactic counterparts, for which the SMI parameter can be consistently derived, as well, from the homogeneous dataset by \citet{snap}. We find that the majority of the considered M31 clusters lie in a significantly different locus, in this plane, with respect to Galactic clusters. In particular, they have redder HB morphologies at a given metallicity, or, in other words, clusters with the same SMI value are $\sim$0.4 dex more metal rich in the Milky Way than in M31. We discuss the possible origin of this difference and we conclude that the most likely explanation is that many globular clusters in the outer halo of M31 formed $\sim$1-2~Gyr later than their counterparts in the outer halo of the Milky Way, while cluster-to-cluster differences in the distribution of He abundance of individual stars may also play a role. It is interesting to note that many of the M31 clusters displaying an HB morphology too red for their metallicity appear to be associated with tidal streams tracing accretion events in the outer halo of the Andromeda galaxy \citep{M10,M12}.  

The results of this analysis are presented and discussed in detail in \citet{sibi}. All the auxiliary data are from the Revised Bologna Catalog of M31 Globular Clusters \citep[RBC][]{gal04,gal06,gal09}\footnote{\tt http://www.bo.astro.it/M31/}. The RBC is constantly updated with new data available from the literature. The last major update occurred in August, 2012.

\begin{acknowledgements}

We acknowledge the financial support of INAF and ASI.
\end{acknowledgements}

\bibliographystyle{aa}

\end{document}